\documentclass[twocolumn,9pt]{extarticle}

\usepackage{amsmath, amsfonts}
\RequirePackage{lmodern}
\usepackage[T1]{fontenc}
\usepackage{graphicx}% Include figure files
\usepackage{dcolumn}% Align table columns on decimal point
\usepackage{bm}% bold math
\usepackage[mathlines]{lineno}% Enable numbering of text and display math
\usepackage[lmargin=48.5pt, rmargin=53.5pt, tmargin=40pt, bmargin=60pt, columnsep=11.5pt]{geometry}
\usepackage{caption}
\usepackage{titling}
\usepackage[superscript, biblabel]{cite}
\usepackage{physics}
\usepackage{setspace}
\usepackage{color}
\usepackage{siunitx}
\usepackage{todonotes}
\usepackage{mhchem}
\usepackage{bm}
\usepackage{url}

\usepackage{etoolbox}
\patchcmd\thebibliography
 {\labelsep}
 {\labelsep\itemsep=-1.9pt}
 {}
 {\typeout{Couldn't patch the command}}
 
\makeatletter
\renewcommand\@cite[2]{%
Ref.~#1\ifthenelse{\boolean{@tempswa}}
{, \nolinebreak[3] #2}{}
}
\renewcommand\@biblabel[1]{#1.}
\makeatother

\newcommand*\samethanks[1][\value{footnote}]{\footnotemark[#1]}

\pretitle{\begin{flushleft}}
\posttitle{\end{flushleft}}
\preauthor{\begin{flushleft}}
\postauthor{\end{flushleft}}

\begin{document}

\title{\textsf{\textbf{{\fontsize{25}{60}\selectfont On-chip integrated laser-driven particle accelerator}}}}

\author{\large\textbf{\textsf{Neil V. Sapra\thanks{Corresponding author: nvsapra@stanford.edu}\hspace{4pt}\thanks{E. L. Ginzton Laboratory, Stanford University, Stanford, CA, USA.}\hspace{3pt}, Ki Youl Yang\samethanks[2]\hspace{3pt}, Dries Vercruysse\samethanks[2]\hspace{3pt}, Kenneth J. Leedle\samethanks[2]\hspace{3pt}, Dylan S. Black\samethanks[2]\hspace{3pt}, R. Joel England\thanks{SLAC National Accelerator Laboratory, Menlo Park, CA, USA.}, Logan Su\samethanks[2]\hspace{3pt}, Yu Miao\samethanks[2]\hspace{3pt}, Olav Solgaard\samethanks[2]\hspace{3pt}, Robert L. Byer\samethanks[2]\hspace{3pt}, Jelena Vu\v{c}kovi\'c\samethanks[2]\vspace{-1.5ex}}}}

\date{\vspace{-4ex}}

\begingroup
\let\center\flushleft
\let\endcenter\endflushleft
\maketitle
\endgroup

\begin{spacing}{0.99}

\noindent\textsf{\textbf{Particle accelerators represent an indispensable tool in science and industry \cite{o2001free,hamm2012industrial,hanna2012rf}. However, the size and cost of conventional radio-frequency accelerators limit the utility and reach of this technology. Dielectric laser accelerators (DLAs) provide a compact and cost-effective solution to this problem by driving accelerator nanostructures with visible or near-infrared (NIR) pulsed lasers, resulting in a $\pmb{10^4}$ reduction of scale \cite{england2014dielectric, wootton2016dielectric}. Current implementations of DLAs rely on free-space lasers directly incident on the accelerating structures, limiting the scalability and integrability of this technology \cite{peralta2013demonstration, breuer2013laser, leedle2015dielectric, leedle2018phase}. Here we present the first experimental demonstration of a waveguide-integrated DLA, designed using a photonic inverse design approach \cite{molesky2018inverse}. These on-chip devices accelerate sub-relativistic electrons of initial energy 83.4 keV by 1.21 keV over \SI{30}{\um}, providing peak acceleration gradients of 40.3 MeV/m. This progress represents a significant step towards a completely integrated MeV-scale dielectric laser accelerator\cite{hughes2018chip}.}}

Dielectric laser accelerators have emerged as a promising alternative to conventional RF accelerators due to the large damage threshold of dielectric materials \cite{stuart1995laser,tien1999short}, the commercial availability of powerful NIR femtosecond pulsed lasers, and the low-cost high-yield nanofabrication processes which produce them. Together, these advantages allow DLAs to make an impact in the development of applications such as tabletop free-electron-lasers, targeted cancer therapies, and compact imaging sources\cite{wootton2016dielectric,plettner2008proposed}.

DLAs are designed by choosing an appropriate pitch and depth of a periodic structure such that the near-fields are phase-matched to electrons of a specific velocity; this scheme is described as the inverse Smith-Purcell effect \cite{plettner2006proposed}. Previous demonstrations of DLAs have relied on free-space lasers directly incident on the accelerating structure, often pillars or gratings made of fused silica or silicon \cite{peralta2013demonstration,breuer2013laser,leedle2015dielectric, chen2018resonant}. Free-space excitation requires bulky optics; therefore, integration with photonic circuits would enable increased scalability, robustness, and impact of this technology.

Integration with photonic waveguides represents a design challenge due to difficulties in accounting for scattering and reflections of the waveguide mode from sub-wavelength features. In this work, we overcome these difficulties through an inverse design approach and develop a waveguide-integrated DLA on a \SI{500}{nm} device layer silicon-on-insulator (SOI) platform. We experimentally demonstrate electron acceleration of \SI{1.21}{keV} over \SI{30}{\um} by coupling light from a pulsed laser, through a broadband grating coupler, and exciting a waveguide mode which acts as the source for the on-chip accelerator (Figure 1a).

\begin{figure*}[th!]
  \vspace{0.655ex}
  \centering
  \includegraphics[width=17cm]{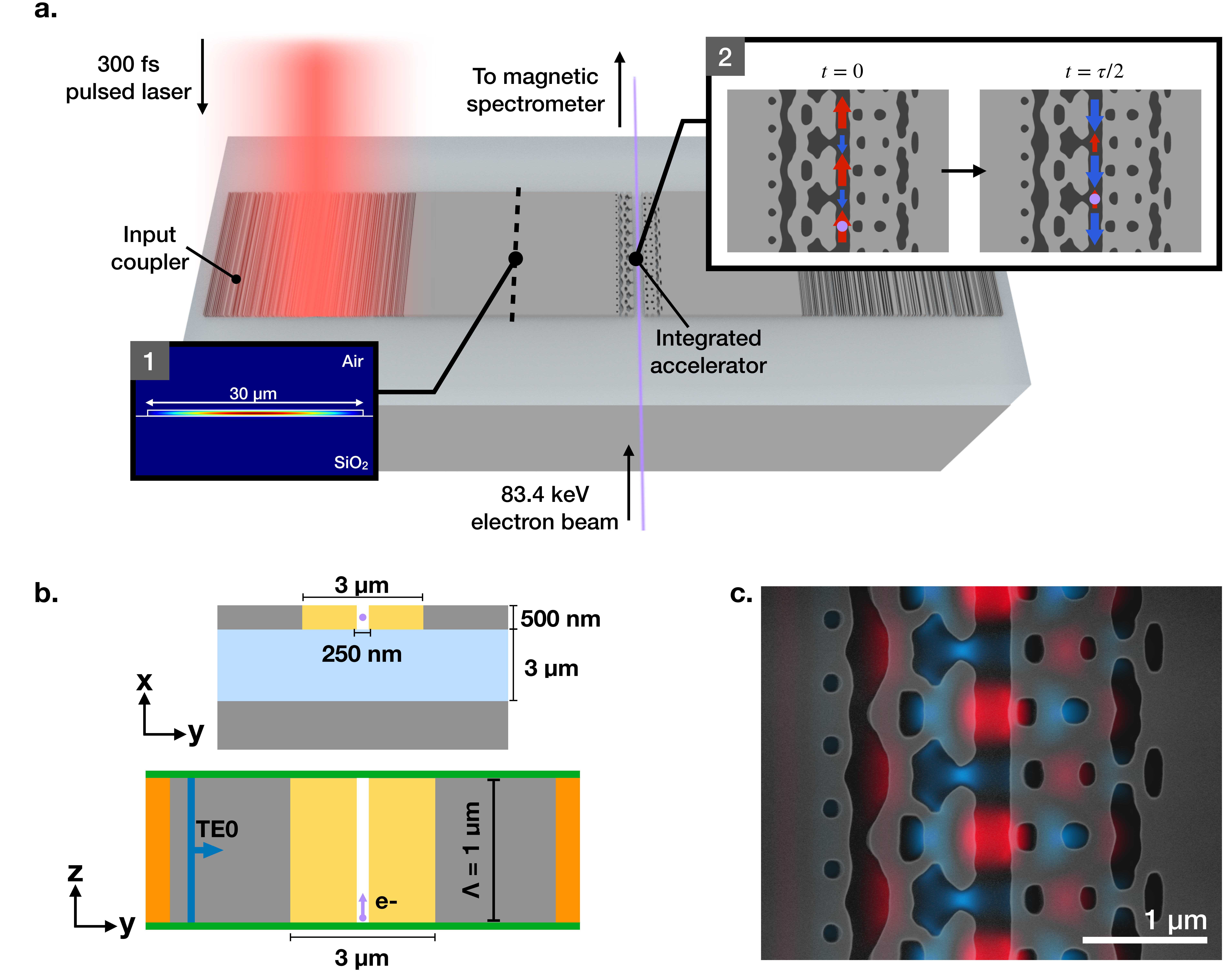}
  \caption{\textbf{Figure 1 \textbar \hspace{0.1pt} Inverse design of on-chip particle accelerator.} \textbf{a.} Schematic depicting components of the on-chip accelerator. An inverse designed grating couples light from a normally incident free-space beam into the fundamental mode of slab-waveguide (Inset 1). The excited waveguide mode then acts as the excitation source for the accelerating structure. The accelerator structure, also designed through inverse design, produces near-fields that are phase-matched to an input electron beam with initial energy \SI{83.4}{keV}. Inset 2 depicts the phase-matched fields and electron at half an optical cycle, $\tau/2$, apart. \textbf{b.} Geometry of the optimization problem. We design on a \SI{500}{nm} silicon (grey), \SI{3}{\um} buried oxide layer (light-blue), silicon-on-insulator (SOI) material stack. Periodic boundary conditions (green) are applied in the z-direction, with a period of $\Gamma = \SI{1}{\um}$, and perfectly matched layers are used in the remaining directions (orange). We optimize the device over a \SI{3}{\um} design region (yellow) with an input source of the fundamental TE0 mode. During the optimization, a \SI{250}{nm} channel for the electron beam to travel in is maintained. \textbf{c.} SEM image of the final accelerator design obtained from the inverse design method. A frame from a time-domain simulation of the accelerating fields, $E_z$, is overlaid.}
  \label{figure:1}
\end{figure*}

To meet the phase-matching condition, the periodicity of the accelerating structure, $\Lambda$, is set by $\Lambda = \beta \lambda$, where $\beta=v/c$ is the ratio of the velocities of the incident electrons to the speed of light, and $\lambda$ is the center frequency of the pump laser\cite{breuer2014dielectric}. To match experimental parameters, we design for a center pump wavelength of \SI{2}{\um} and an input electron velocity of $v = 0.5c$, resulting in an accelerator period of $\Lambda = \SI{1}{\um}$. Figure 1b captures the geometry of the optimization problem. Using an in-house inverse design software suite developed by our team, SPINS \cite{spins,piggott2015inverse, piggott2017fabrication, su2017inverse}, we optimize the design of the accelerator over a \SI{3}{\um} region, ensuring to preserve a \SI{250}{nm} center channel for electron propagation. The accelerator is simulated with a fully-3D finite-difference frequency-domain (FDFD) solver, meshed with a uniform grid of spacing \SI{30}{nm}. Periodic boundary conditions are applied in the direction of electron propagation (z-axis) to enforce the accelerator period, and perfectly matched layers are used in the remaining axes \cite{shin2012choice}. The structure is excited with the fundamental slab waveguide mode, and the following 3D optimization problem is solved:
 
\begin{equation}
\label{eqn:invdes}
\begin{aligned}
& \underset{p, E^1, E^2, \dots, E^m}{\text{maximize}}
& & \sum_{i=1}^m |G_z(E^i)| - |G_y(E^i)| \\
& \text{subject to}
& & \nabla \times \frac{1}{\mu_0} \nabla \times E^i - \omega_i^2 \epsilon(p)E^i   = -i\omega_i J_i, \\
& & & i = 1, 2, \dots, m
\end{aligned}
\end{equation}

We express the acceleration gradient, $G_z(E^i)$, the integrated field the electron experiences as it travels through one period of the accelerator, in the frequency domain \cite{hughes2017method}. The second term, $G_y(E^i)$, corresponds to the deflecting transverse gradients, which we penalize. The fields are subject to Maxwell's Equations, and the permittivity of the device, $\epsilon(p)$, is parameterized by a vector of design variables, $p$. In order to have good spectral overlap with the broadband input pulsed laser spectrum, each objective function evaluation is the sum of $m=3$ simulations, each with a different input source frequency, $\omega_i$. The three simulations sample a \SI{30}{nm} total bandwidth around \SI{2}{\um}. During the final optimization stage, an additional constraint is introduced to enforce a minimum fabricable feature size of \SI{80}{nm}. Further details regarding the design of the accelerator can be found in the Methods section. A scanning electron microscope (SEM) image of a fabricated optimized accelerator is shown in Figure 1c, with a frame from simulated time-domain fields overlaid.

As the optimization was carried out with periodic boundary conditions, we verified the performance of a finite-length 30 period accelerator structure in a 3D finite-difference time-domain (FDTD) simulation \cite{lumericalsolution}. In order to determine the peak operating wavelength of the accelerator, we compute the frequency domain gradients (Supplemental Information), obtaining the acceleration gradient spectrum. As seen in Figure 2a, the spectrum peaks at \SI{1.964}{\um}, indicating a shift from the design wavelength due to the finite-length and numerical dispersion\cite{taflove2005computational}. Furthermore, we see the broadband optimization successfully produces a structure with good spectral overlap with the \SI{300}{fs} input pulse ($\Delta\lambda = \SI{18.9}{nm}$ FWHM centered at \SI{1.964}{\um}). With knowledge of the peak operating wavelength, we model experimental conditions by injecting a \SI{300}{fs} fundamental mode source centered at \SI{1.964}{\um} into the finite-length structure. We compute the time-domain acceleration gradients ($G_z$) and deflecting gradients ($G_y$) of the resulting simulation through:

\begin{equation}
\label{eqn:time_domain_gradient}
\begin{aligned}
G_k(t_0) = \frac{1}{L}\int_{0}^{L}{E_k(z,t_0+z/\beta c_0)}dz,
\end{aligned}
\end{equation}

\noindent where $t_0$ is the delay between the time of source injection and the electron entering the accelerator channel, and $L = \SI{30}{\um}$ is the length of the accelerator. At optimal time delay $t_0$ (see Supplemental Information for determination of $t_0$), the accelerating and deflecting gradients are seen in Figure 2b, with normalization such that the peak magnitude of fields in the waveguide mode is unity. We see that at $\beta = 0.51$, the deflecting gradients approach zero, indicating the efficacy of the penalty term on the transverse gradients during optimization.

\begin{figure}[h]
  \vspace{0.655ex}
  \includegraphics[width=8.8cm]{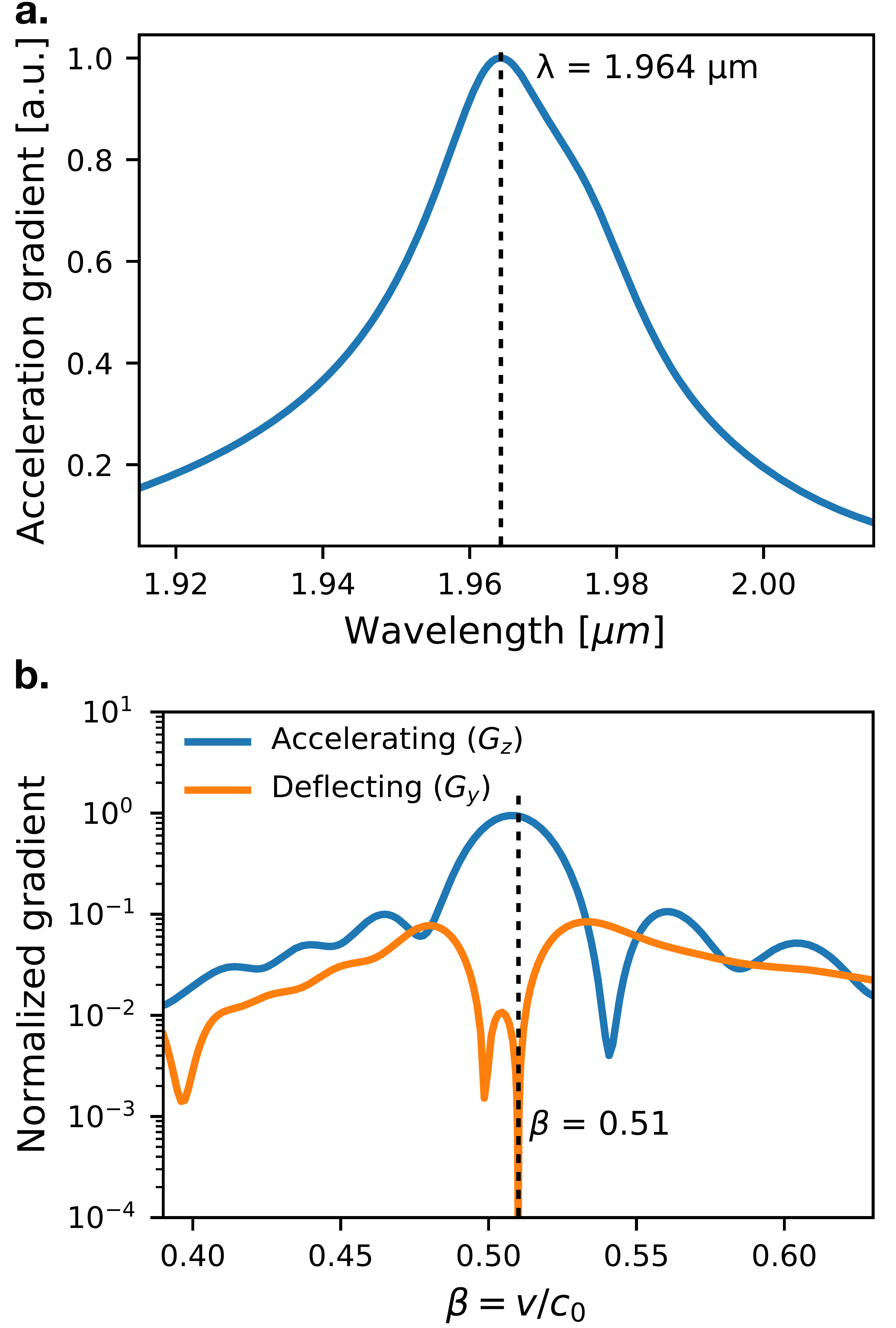}
  \caption{\textbf{Figure 2 \textbar \hspace{0.1pt} Simulated performance of optimized accelerators.} \textbf{a.} Acceleration gradient spectrum for a finite-length accelerator composed of 30 periods. \textbf{b.} Accelerating gradients and transverse deflecting gradients as a function of input electron velocity from simulated time-domain fields. At $\beta = 0.51$ (\SI{83.4}{keV}), the deflecting gradients tend toward zero while the accelerating gradients are near-maximum. Fields normalized such that the peak electric field in the waveguide mode is unity.}
  \label{figure:2}
\end{figure}

We fabricated a 30-period accelerator, waveguides, and grating couplers on a \SI{500}{nm} thick SOI wafer using electron beam lithography and reactive ion etching. To provide clearance for the electron beam, the area surrounding the accelerator is etched with an additional photolithography step to form a "mesa." Complete fabrication details can be found in the Methods section. Figure 3 shows an SEM image of the fabricated single-stage accelerator on top of a mesa.

\begin{figure}[h]
  \vspace{0.655ex}
  \includegraphics[width=8.8cm]{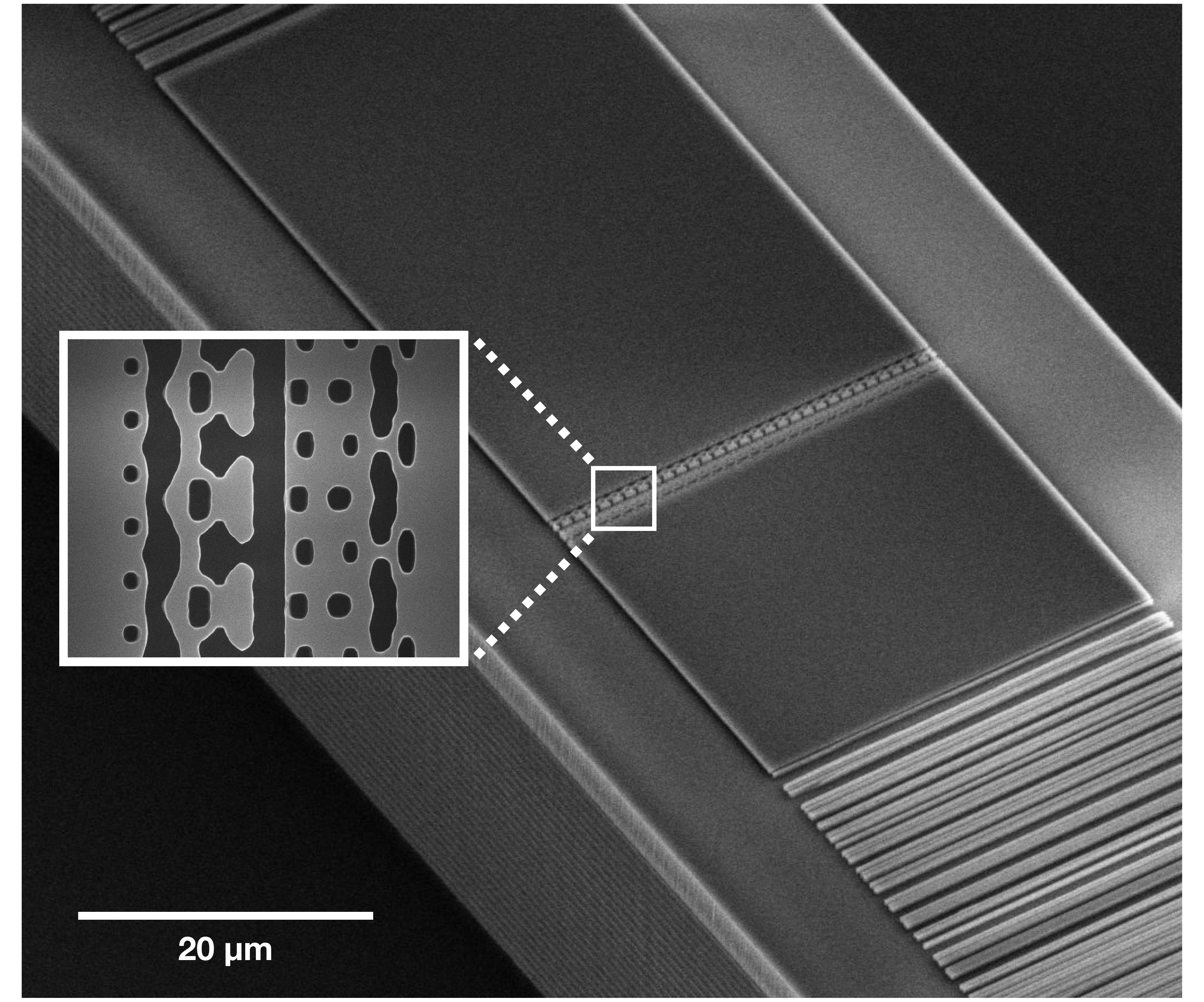}
  \caption{\textbf{Figure 3 \textbar \hspace{0.1pt} Fabricated single-stage accelerator.} SEM image of a single-stage accelerator of 30 periods fabricated on a \SI{500}{nm} SOI stack. The accelerator sits on a \SI{25}{\um} tall mesa structure to provide clearance for the input electron beam.}
  \label{figure:3}
\end{figure}

The experimental setup is adapted from previous direct-incidence pillar experiments to support normal incidence on a grating coupler \cite{leedle2018phase,black2019laser}. Light generated from a \SI{300}{fs} FWHM pulse-length, \SI{100}{kHz} repetition rate, optical parametric amplifier (OPA) is focused to a \SI{40}{\um} $1/e^2$ diameter beam, normally incident on the input grating coupler to excite the fundamental waveguide mode of the slab-waveguide (see Methods section for grating coupler design). A custom-built scanning transmission electron microscope is used as the source for the electron beam which travels through the channel in the accelerator structure with an initial energy of \SI{83.4}{keV} ($v = 0.51c$). Electrons which pass through the accelerator are separated by energy in a magnetic spectrometer, before terminating at a micro-channel plate (MCP) detector to image the energy distribution (see Methods section for additional detail).

The electron energy spectra in Figure 4a clearly show that electrons have been successfully accelerated by our structure. The blue curve depicts the energy spectrum of the electrons passing through the accelerator structure with the laser off, and the red curve shows the energy spectrum when the laser (\SI{3}{mW} average power, \SI{335}{MV/m} peak field, at \SI{1940}{nm}) is incident on the grating coupler. As the bunch-length is larger than the optical cycle, we observe symmetric broadening of the energy spectrum, consisting of two populations of accelerated and decelerated electrons. The energy corresponding to the centroid of the accelerated population is often referred to as the "shoulder energy." \cite{leedle2015dielectric} We determine this value by locating the energy at which there is the greatest linear difference between the laser-on and laser-off curves; this definition provides a metric that is consistent with shoulder energies as determined in previous DLA experiments \cite{leedle2015dielectric, leedle2018phase}. For the spectra shown in Figure 4a, the shoulder energy is \SI{83.77}{keV}, indicating a shoulder energy modulation of \SI{0.33}{keV}. In addition, the maximal energy detected is \SI{84.65}{keV}, corresponding to an maximum energy gain of \SI{1.21}{keV}. The dotted red curve depicts simulated performance of the accelerator based on particle tracking simulations (see Methods section), providing good agreement with the experimental spectrum.

\begin{figure*}[th!]
  \vspace{0.655ex}
  \centering
  \includegraphics[width=17cm]{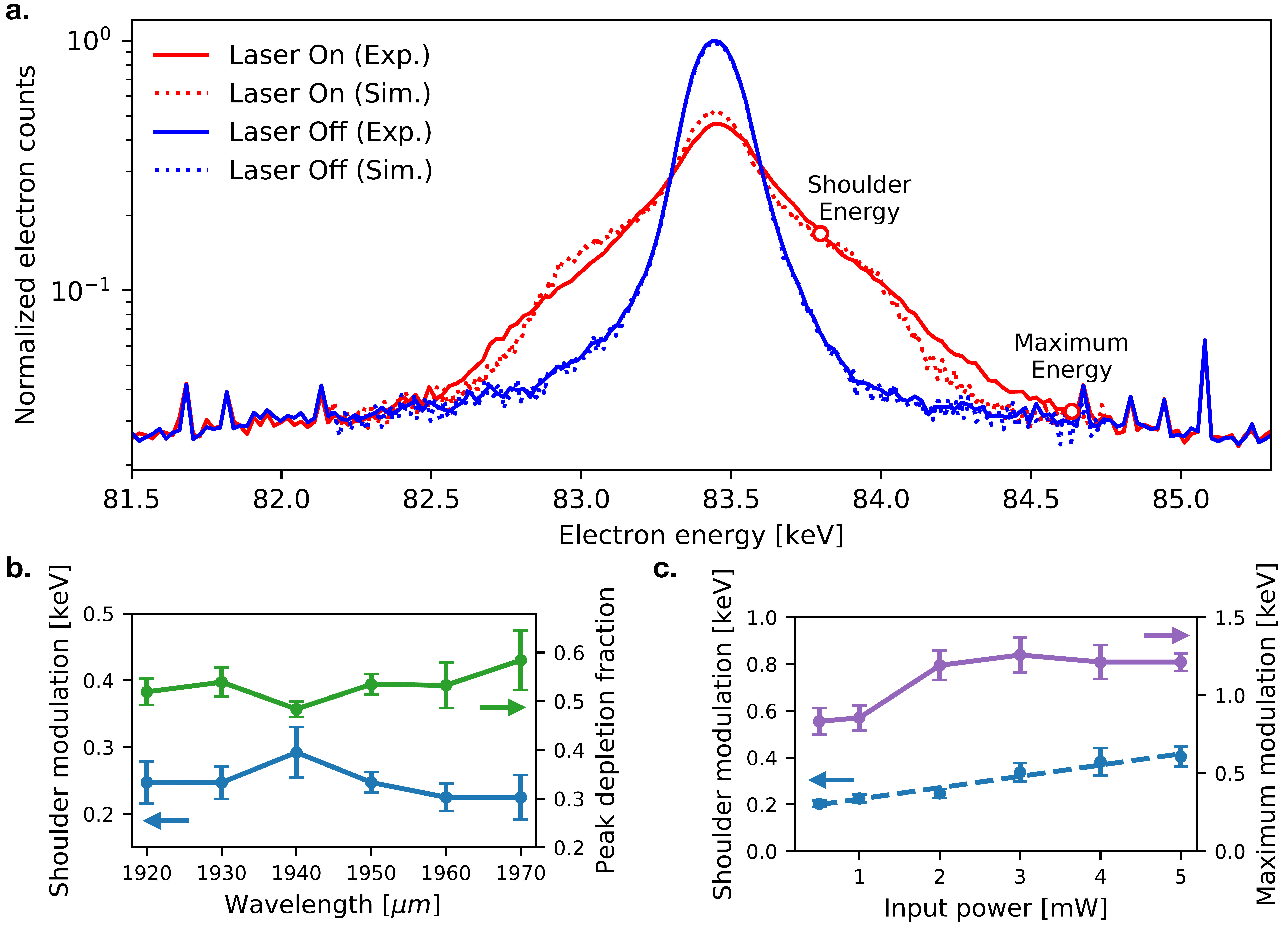}
  \caption{\textbf{Figure 4 \textbar \hspace{0.1pt} Experimental verification of acceleration.} \textbf{a.} Electron energy spectrum without laser incident (blue curve) and with laser, \SI{3.0}{mW}, \SI{335}{MV/m} peak field, at $\lambda = \SI{1940}{nm}$, incident (red curve) on the grating coupler. Open red circles denote the location of the shoulder energy modulation, and maximum energy modulation. Simulated spectrum based on particle tracking simulations shown in dotted red curve. \textbf{b.} Shoulder energy modulation (blue, left axis) and peak depletion (green, right axis) for a fixed power at \SI{2.75}{mW}, \SI{321}{MV/m} peak field, as a function of varying the wavelength of the pump laser. \textbf{c.} Shoulder energy modulation (blue, left axis) and maximum energy modulation (purple, right axis) at fixed wavelength of \SI{1940}{nm}, as a function of input power.}
  \label{figure:4}
\end{figure*}

To determine the peak operating wavelength of our accelerator, we fix the average power of the incident laser pulses to be \SI{2.75}{mW} (\SI{321}{MV/m} peak field) and sweep the wavelength (Figure 4b). We observe a peak in the shoulder energy modulation at \SI{1940}{nm}. Moreover, the ratio of laser-on to laser-off counts at the center energy, referred to as the "peak depletion," is optimal at \SI{1940}{nm}, indicating the greatest population of modulated electrons at this wavelength. Therefore, the shoulder energy modulation and peak depletion suggest an operating wavelength of \SI{1940}{nm}. 

Fixing the wavelength at \SI{1940}{nm} (Figure 4c), we conduct a sweep of the input power from \SI{0.5}{mW} to \SI{5}{mW} (\SI{137}{MV/m} to \SI{433}{MV/m} peak fields). We observe linear behavior in the shoulder energy modulation as the input power is increased. In the maximum modulation, we observe saturation beyond \SI{2}{mW} (\SI{274}{MV/m} peak field), indicating damage to the input couplers beyond this point. At the maximal power, we obtain the largest shoulder energy modulation of \SI{0.4}{keV}.

In this letter, we have designed and experimentally verified the first waveguide-integrated DLA structure. The design of this structure was made possible through the use of photonics inverse design methodologies developed by our team members. The fabricated and experimentally demonstrated devices accelerate electrons of an initial energy of \SI{83.4}{keV} by a maximum energy gain of \SI{1.21}{keV} over \SI{30}{\um}, demonstrating acceleration gradients of \SI{40.3}{MeV/m}. In this integrated form, these devices can be cascaded to reach MeV-scale energies, capitalizing on the inherent scalability of photonic circuits \cite{hughes2018chip}. To this end, future work will focus on multi-stage demonstrations, as well as exploring new design and material solutions to obtain larger gradients. The work shown here paves the way toward compact accelerator technologies for the development of applications in industrial processing, materials research, and medicine. 

\end{spacing}

% \vspace{-1.7ex}
\newpage

\section*{{\fontsize{11}{11}\textsf{\textbf{Methods\vspace{-1.2 ex}}}}}
{\footnotesize
\begin{spacing}{1.05}

\textbf{Inverse design of on-chip accelerator} The core of the inverse design algorithm used in this work is identical to the process described in previous work \cite{piggott2015inverse, piggott2017fabrication, su2017inverse}. The major departure is the objective function used. The objective function which we maximize, for a single input mode at frequency $\omega$, is written as:

$$F = |G_z(E)| - |G_y(E)|,$$

\noindent where $G_z(E)$ represents the longitudinal acceleration gradient and $G_y(E)$ represents the transverse deflecting gradient, as computed in the frequency domain\cite{hughes2017method}. More explicitly, the form of each of these terms is:

\begin{equation*}
\begin{split}
    G_z(E) = \alpha_1\int_{0}^{L}\int_{-g/2}^{0}e^{-i \omega z/\beta c_0}E_z(z,\omega)dy\,dz \\
    + \alpha_2\int_{0}^{L}\int_{0}^{g/2}e^{-i \omega z/\beta c_0}E_z(z,\omega)dy\,dz,
\end{split}
\end{equation*}

\noindent and

\begin{equation*}
\begin{split}
    G_y(E) = \alpha_3\int_{0}^{L}\int_{-g/2}^{g/2}e^{-i \omega z/\beta c_0}E_y(z,\omega)dy\,dz.
\end{split}
\end{equation*}

The two terms in $G_z(E)$ are responsible for the acceleration gradient. The acceleration gradient is computed across the entire electron propagation channel of length $L$ and width $g$ for an electron travelling at velocity $v_e=\beta c_0$. Note, the integral is decomposed into the left and right halves.  As the structure is driven from a single-side, the accelerating fields tend to decay from the side of illumination (left of the accelerator channel), resulting in non-uniform acceleration of the electron beam across the gap. Therefore, to promote symmetric fields, we add a weight to the fields in the right-half of the accelerator channel  ($\alpha_1 < \alpha_2$). The end result of this weighting is to produce accelerating fields with greater symmetry across the gap, as illustrated in Figure 1c.

The second term in the objective function, $G_y(E)$, is of similar form as the previous; however, we evaluate the transverse gradients from the deflecting $E_y$ fields. This integral is weighted equally across the gap by the coefficient $\alpha_3$.

\textbf{Fabrication of accelerator} The devices were fabricated at the Stanford Nanofabrication Facility (SNF) and Stanford Nanofabrication Shared Facilities (SNSF) on \SI{500}{nm} device layer, \SI{3}{\um} buried oxide layer silicon-on-insulator pieces. A JEOL JBC-6300FS electron-beam lithography tool was used to pattern \SI{330}{nm} thick ZEP-520A electron-beam resist with the accelerator, waveguides, and grating couplers. After development, the samples were etched using reactive-ion etching with a \ce{C2F6} breakthrough step, and \ce{BCl3}/\ce{Cl2}/\ce{O2} chemistry main etch. Resist was removed using an overnight soak in 1165, followed by a Piranha clean, 4:1 ratio of sulphuric acid and 30\% hydrogen peroxide.

The mesa structure, the platform to provide clearance for the input electron beam, was defined using contact photolithography. The cleaned, patterned samples were dehydrated and primed with HMDS in a YES Oven. SPR220-3 photoresist was spun at 2000 rpm for a nominal thickness of \SI{4}{\um}. The samples were exposed with a Karl Suss MA-6 Contact Aligner with a 4 second, \SI{365}{nm}, $\SI{15}{mW/cm^2}$ exposure. Exposure was preceded and followed by a $115^{\circ}$ C bake for 90 seconds. The photolithography was developed with 60 seconds of agitation in MF-26A developer, followed by agitation in water. The mesa was etched to be roughly \SI{25}{\um} deep through three separate plasma etches: another \ce{BCl3}/\ce{Cl2}/\ce{O2} based etch for the silicon layer, \ce{CF4}/\ce{CHF3} etch for the buried oxide layer, and the Bosch process for a deep-silicon-etch into the silicon substrate. Lastly, the sample was cleaned using another overnight 1165 soak, followed by Piranha bath.

\textbf{Grating coupler design} The waveguide mode which acts as the source to the accelerator is excited through a grating coupler. The couplers are required to be broadband in order to couple the bandwidth of the input pulsed laser, as well as to account for spectral drift due to fabrication imperfections in the grating coupler or accelerator. Furthermore, the couplers must reject coupling into the higher order modes of the \SI{500}{nm} thick by \SI{30}{\um} wide waveguide. To meet both of these specifications, we again make use of the inverse design method to design the grating coupler, as detailed in previous work \cite{su2018fully, sapra2019inverse}. The resulting optimized couplers with a conservative minimum feature size of \SI{120}{nm} couple normally incident light with a spot size of \SI{40}{\um} with a simulated (3D, FDTD) coupling efficiency into the fundamental TE0 waveguide mode of 14.3\% at \SI{1.964}{\um}, and a \SI{1}{dB} bandwidth of \SI{45}{nm}. 95\% of the total transmission is into the fundamental mode, with the remaining coupled to the higher order modes (Supplemental Information). Higher coupling efficiency grating couplers can be designed at the cost of bandwidth and smaller feature sizes; however, bandwidth and robustness were prioritized for this experiment.

\textbf{Measurement of electron energy modulation} 
A custom-built scanning transmission electron microscope is used as the source for the electron beam. Ultraviolet pulses ($\SI{300}{fs}\pm\SI{10}{nm}$ FWHM, \SI{100}{kHz}, \SI{256}{nm}) illuminate a flat copper photo-cathode to produce a \SI{2.2}{mrad} full-angle divergence electron beam at \SI{83.4}{keV} ($v = 0.51c$). A solenoid lens focuses the electron beam to \SI{490}{nm} diameter spot, as measured by a knife-edge scan at the location of the accelerator. The Rayleigh length of the electron beam is approximately \SI{150}{\um}, resulting in a near-uniform spot-size over the \SI{30}{\um} length of the accelerator. The light source is a $\SI{300}{fs}\pm\SI{10}{nm}$ FWHM pulse-length, \SI{100}{kHz} repetition rate, optical parametric amplifier (OPA), focused to beam with $1/e^2$ diameter of $40 \pm \SI{4}{\um}$, normally incident on the input grating coupler. Alignment of the beam to the grating coupler is done through imaging the back-reflection of the laser which provides the ability to discern features of the structure. Electrons which pass through the accelerator enter a magnetic spectrometer with an energy resolution of \SI{40}{eV} with a FWHM of \SI{250}{eV}. The energy-separated electrons finally terminate at a micro-channel plate (MCP) detector for imaging. The electron energy spectra are obtained by averaging at least five frames, with each frame consisting of two seconds of integration on the MCP detector (see Supplemental Information for frame data). From these spectra, the shoulder energy was defined to be the energy at which there is the greatest linear difference between the laser-on and laser-off curves. The maximum energy was defined to be the trailing energy at which the laser-on and laser-off curves intersect. Error estimation in Figure 4b and 4c was obtained by computing the standard deviation of the quantities of the individual frames which comprised the average spectrum.

\textbf{Particle tracking simulation} The particle tracking code General Particle Tracer (GPT) was used to calculate the laser-on electron spectrum (red dashed curve in Figure 4a). A three-dimensional map of the complex fields, based on the simulations in Figure 1c, was generated and imported into GPT, and a realistic electron beam distribution was propagated through the accelerating channel, with RMS dimensions $\sigma_x = \sigma_y$ = 120 nm and normalized transverse emittances $\epsilon_{nx,y}$ = 6 pm-rad. Space charge effects were turned off to be consistent with the low beam current in the experiment. The simulated initial electron spectrum in Figure 4a (dashed blue curve) is a Lorentzian fit to the experimental laser-off data (solid blue curve). The field map region was matched to the dimensions of the accelerating channel of the DLA (0.5 $\times$ 0.25 $\times$ 30 $\mu$m) and a collimating filter was used to remove particles whose trajectories intersected the walls. Since the code does not allow for a time-varying field amplitude, the temporal envelope of the laser was included by summing multiple spectra corresponding to successive slices of the electron beam, with the average gradient $G_k$ of each spectrum matched to Eq. \ref{eqn:time_domain_gradient} for that slice. The laser field was modelled as a Gaussian $E_k(z,t) \propto \exp (-t^2/\tau^2-z^2/w_0^2)$ where $\tau$ = 272 fs, $w_0$ = 12.7 $\mu$m are the dispersion-corrected pulse duration and transverse mode size in the waveguide respectively. The electron temporal profile was a Gaussian with FWHM of 500 fs, consistent with the experimental laser-electron correlation function. For optimal matching of the width of the spectrum to the data, the simulated laser-on curve corresponds to a peak average gradient $G_k$ = 48.4 MeV/m, but the highest energy particles fall below the experimental noise floor. Consequently, the experimentally derived gradient of 40.3 MeV/m is likely a conservative estimate.
\end{spacing}
}

\vspace{-0.9ex}

\bibliographystyle{naturemag}
{\footnotesize\bibliography{bibliography}}% Produces the bibliography via BibTeX.

\begin{thebibliography}{10}
\expandafter\ifx\csname url\endcsname\relax
  \def\url#1{\texttt{#1}}\fi
\expandafter\ifx\csname urlprefix\endcsname\relax\def\urlprefix{URL }\fi
\providecommand{\bibinfo}[2]{#2}
\providecommand{\eprint}[2][]{\url{#2}}

\bibitem{o2001free}
\bibinfo{author}{O'shea, P.~G.} \& \bibinfo{author}{Freund, H.~P.}
\newblock \bibinfo{title}{Free-electron lasers: status and applications}.
\newblock \emph{\bibinfo{journal}{Science}} \textbf{\bibinfo{volume}{292}},
  \bibinfo{pages}{1853--1858} (\bibinfo{year}{2001}).

\bibitem{hamm2012industrial}
\bibinfo{author}{Hamm, R.~W.} \& \bibinfo{author}{Hamm, M.~E.}
\newblock \emph{\bibinfo{title}{Industrial accelerators and their
  applications}} (\bibinfo{publisher}{World Scientific}, \bibinfo{year}{2012}).

\bibitem{hanna2012rf}
\bibinfo{author}{Hanna, S.}
\newblock \emph{\bibinfo{title}{RF linear accelerators for medical and
  industrial applications}} (\bibinfo{publisher}{Artech House},
  \bibinfo{year}{2012}).

\bibitem{england2014dielectric}
\bibinfo{author}{England, R.~J.} \emph{et~al.}
\newblock \bibinfo{title}{Dielectric laser accelerators}.
\newblock \emph{\bibinfo{journal}{Reviews of Modern Physics}}
  \textbf{\bibinfo{volume}{86}}, \bibinfo{pages}{1337} (\bibinfo{year}{2014}).

\bibitem{wootton2016dielectric}
\bibinfo{author}{Wootton, K.}, \bibinfo{author}{McNeur, J.} \&
  \bibinfo{author}{Leedle, K.}
\newblock \bibinfo{title}{Dielectric laser accelerators: designs, experiments,
  and applications}.
\newblock In \emph{\bibinfo{booktitle}{Reviews of Accelerator Science and
  Technology: Volume 9: Technology and Applications of Advanced Accelerator
  Concepts}}, \bibinfo{pages}{105--126} (\bibinfo{publisher}{World Scientific},
  \bibinfo{year}{2016}).

\bibitem{peralta2013demonstration}
\bibinfo{author}{Peralta, E.} \emph{et~al.}
\newblock \bibinfo{title}{Demonstration of electron acceleration in a
  laser-driven dielectric microstructure}.
\newblock \emph{\bibinfo{journal}{Nature}} \textbf{\bibinfo{volume}{503}},
  \bibinfo{pages}{91} (\bibinfo{year}{2013}).

\bibitem{breuer2013laser}
\bibinfo{author}{Breuer, J.} \& \bibinfo{author}{Hommelhoff, P.}
\newblock \bibinfo{title}{Laser-based acceleration of nonrelativistic electrons
  at a dielectric structure}.
\newblock \emph{\bibinfo{journal}{Physical review letters}}
  \textbf{\bibinfo{volume}{111}}, \bibinfo{pages}{134803}
  (\bibinfo{year}{2013}).

\bibitem{leedle2015dielectric}
\bibinfo{author}{Leedle, K.~J.} \emph{et~al.}
\newblock \bibinfo{title}{Dielectric laser acceleration of sub-100 kev
  electrons with silicon dual-pillar grating structures}.
\newblock \emph{\bibinfo{journal}{Optics letters}}
  \textbf{\bibinfo{volume}{40}}, \bibinfo{pages}{4344--4347}
  (\bibinfo{year}{2015}).

\bibitem{leedle2018phase}
\bibinfo{author}{Leedle, K.~J.} \emph{et~al.}
\newblock \bibinfo{title}{Phase-dependent laser acceleration of electrons with
  symmetrically driven silicon dual pillar gratings}.
\newblock \emph{\bibinfo{journal}{Optics letters}}
  \textbf{\bibinfo{volume}{43}}, \bibinfo{pages}{2181--2184}
  (\bibinfo{year}{2018}).

\bibitem{molesky2018inverse}
\bibinfo{author}{Molesky, S.} \emph{et~al.}
\newblock \bibinfo{title}{Inverse design in nanophotonics}.
\newblock \emph{\bibinfo{journal}{Nature Photonics}}
  \textbf{\bibinfo{volume}{12}}, \bibinfo{pages}{659} (\bibinfo{year}{2018}).

\bibitem{hughes2018chip}
\bibinfo{author}{Hughes, T.~W.} \emph{et~al.}
\newblock \bibinfo{title}{On-chip laser-power delivery system for dielectric
  laser accelerators}.
\newblock \emph{\bibinfo{journal}{Physical Review Applied}}
  \textbf{\bibinfo{volume}{9}}, \bibinfo{pages}{054017} (\bibinfo{year}{2018}).

\bibitem{stuart1995laser}
\bibinfo{author}{Stuart, B.}, \bibinfo{author}{Feit, M.},
  \bibinfo{author}{Rubenchik, A.}, \bibinfo{author}{Shore, B.} \&
  \bibinfo{author}{Perry, M.}
\newblock \bibinfo{title}{Laser-induced damage in dielectrics with nanosecond
  to subpicosecond pulses}.
\newblock \emph{\bibinfo{journal}{Physical review letters}}
  \textbf{\bibinfo{volume}{74}}, \bibinfo{pages}{2248} (\bibinfo{year}{1995}).

\bibitem{tien1999short}
\bibinfo{author}{Tien, A.-C.}, \bibinfo{author}{Backus, S.},
  \bibinfo{author}{Kapteyn, H.}, \bibinfo{author}{Murnane, M.} \&
  \bibinfo{author}{Mourou, G.}
\newblock \bibinfo{title}{Short-pulse laser damage in transparent materials as
  a function of pulse duration}.
\newblock \emph{\bibinfo{journal}{Physical Review Letters}}
  \textbf{\bibinfo{volume}{82}}, \bibinfo{pages}{3883} (\bibinfo{year}{1999}).

\bibitem{plettner2008proposed}
\bibinfo{author}{Plettner, T.} \& \bibinfo{author}{Byer, R.}
\newblock \bibinfo{title}{Proposed dielectric-based microstructure laser-driven
  undulator}.
\newblock \emph{\bibinfo{journal}{Physical Review Special Topics-Accelerators
  and Beams}} \textbf{\bibinfo{volume}{11}}, \bibinfo{pages}{030704}
  (\bibinfo{year}{2008}).

\bibitem{plettner2006proposed}
\bibinfo{author}{Plettner, T.}, \bibinfo{author}{Lu, P.} \&
  \bibinfo{author}{Byer, R.}
\newblock \bibinfo{title}{Proposed few-optical cycle laser-driven particle
  accelerator structure}.
\newblock \emph{\bibinfo{journal}{Physical Review Special Topics-Accelerators
  and Beams}} \textbf{\bibinfo{volume}{9}}, \bibinfo{pages}{111301}
  (\bibinfo{year}{2006}).

\bibitem{chen2018resonant}
\bibinfo{author}{Chen, Z.}, \bibinfo{author}{Koyama, K.},
  \bibinfo{author}{Uesaka, M.}, \bibinfo{author}{Yoshida, M.} \&
  \bibinfo{author}{Zhang, R.}
\newblock \bibinfo{title}{Resonant enhancement of accelerating gradient with
  silicon dual-grating structure for dielectric laser acceleration of
  subrelativistic electrons}.
\newblock \emph{\bibinfo{journal}{Applied Physics Letters}}
  \textbf{\bibinfo{volume}{112}}, \bibinfo{pages}{034102}
  (\bibinfo{year}{2018}).

\bibitem{breuer2014dielectric}
\bibinfo{author}{Breuer, J.}, \bibinfo{author}{Graf, R.},
  \bibinfo{author}{Apolonski, A.} \& \bibinfo{author}{Hommelhoff, P.}
\newblock \bibinfo{title}{Dielectric laser acceleration of nonrelativistic
  electrons at a single fused silica grating structure: Experimental part}.
\newblock \emph{\bibinfo{journal}{Physical Review Special Topics-Accelerators
  and Beams}} \textbf{\bibinfo{volume}{17}}, \bibinfo{pages}{021301}
  (\bibinfo{year}{2014}).

\bibitem{spins}
\bibinfo{author}{Vuckovic, J.} \emph{et~al.}
\newblock \bibinfo{title}{Inverse design software for nanophotonic structures -
  {S}pins}.
\newblock \bibinfo{howpublished}{\url{https://github.com/stanfordnqp/spins-b}}
  (\bibinfo{year}{2018}).
\newblock \bibinfo{note}{Stanford OTL Reference S18-012}.

\bibitem{piggott2015inverse}
\bibinfo{author}{Piggott, A.~Y.} \emph{et~al.}
\newblock \bibinfo{title}{Inverse design and demonstration of a compact and
  broadband on-chip wavelength demultiplexer}.
\newblock \emph{\bibinfo{journal}{Nature Photonics}}
  \textbf{\bibinfo{volume}{9}}, \bibinfo{pages}{374} (\bibinfo{year}{2015}).

\bibitem{piggott2017fabrication}
\bibinfo{author}{Piggott, A.~Y.}, \bibinfo{author}{Petykiewicz, J.},
  \bibinfo{author}{Su, L.} \& \bibinfo{author}{Vu{\v{c}}kovi{\'c}, J.}
\newblock \bibinfo{title}{Fabrication-constrained nanophotonic inverse design}.
\newblock \emph{\bibinfo{journal}{Scientific reports}}
  \textbf{\bibinfo{volume}{7}}, \bibinfo{pages}{1786} (\bibinfo{year}{2017}).

\bibitem{su2017inverse}
\bibinfo{author}{Su, L.}, \bibinfo{author}{Piggott, A.~Y.},
  \bibinfo{author}{Sapra, N.~V.}, \bibinfo{author}{Petykiewicz, J.} \&
  \bibinfo{author}{Vuckovic, J.}
\newblock \bibinfo{title}{Inverse design and demonstration of a compact on-chip
  narrowband three-channel wavelength demultiplexer}.
\newblock \emph{\bibinfo{journal}{Acs Photonics}} \textbf{\bibinfo{volume}{5}},
  \bibinfo{pages}{301--305} (\bibinfo{year}{2017}).

\bibitem{shin2012choice}
\bibinfo{author}{Shin, W.} \& \bibinfo{author}{Fan, S.}
\newblock \bibinfo{title}{Choice of the perfectly matched layer boundary
  condition for frequency-domain maxwell's equations solvers}.
\newblock \emph{\bibinfo{journal}{Journal of Computational Physics}}
  \textbf{\bibinfo{volume}{231}}, \bibinfo{pages}{3406--3431}
  (\bibinfo{year}{2012}).

\bibitem{hughes2017method}
\bibinfo{author}{Hughes, T.}, \bibinfo{author}{Veronis, G.},
  \bibinfo{author}{Wootton, K.~P.}, \bibinfo{author}{England, R.~J.} \&
  \bibinfo{author}{Fan, S.}
\newblock \bibinfo{title}{Method for computationally efficient design of
  dielectric laser accelerator structures}.
\newblock \emph{\bibinfo{journal}{Optics express}}
  \textbf{\bibinfo{volume}{25}}, \bibinfo{pages}{15414--15427}
  (\bibinfo{year}{2017}).

\bibitem{lumericalsolution}
\bibinfo{title}{Lumerical {I}nc.}
\newblock \urlprefix\url{https://www.lumerical.com/tcad-products/fdtd/}.

\bibitem{taflove2005computational}
\bibinfo{author}{Taflove, A.} \& \bibinfo{author}{Hagness, S.~C.}
\newblock \emph{\bibinfo{title}{Computational electrodynamics: the
  finite-difference time-domain method}} (\bibinfo{publisher}{Artech house},
  \bibinfo{year}{2005}).

\bibitem{black2019laser}
\bibinfo{author}{Black, D.~S.} \emph{et~al.}
\newblock \bibinfo{title}{Laser-driven electron lensing in silicon
  microstructures}.
\newblock \emph{\bibinfo{journal}{Physical Review Letters}}
  \textbf{\bibinfo{volume}{122}}, \bibinfo{pages}{104801}
  (\bibinfo{year}{2019}).

\bibitem{su2018fully}
\bibinfo{author}{Su, L.} \emph{et~al.}
\newblock \bibinfo{title}{Fully-automated optimization of grating couplers}.
\newblock \emph{\bibinfo{journal}{Optics express}}
  \textbf{\bibinfo{volume}{26}}, \bibinfo{pages}{4023--4034}
  (\bibinfo{year}{2018}).

\bibitem{sapra2019inverse}
\bibinfo{author}{Sapra, N.~V.} \emph{et~al.}
\newblock \bibinfo{title}{Inverse design and demonstration of broadband grating
  couplers}.
\newblock \emph{\bibinfo{journal}{IEEE Journal of Selected Topics in Quantum
  Electronics}}  (\bibinfo{year}{2019}).

\end{thebibliography}

\vspace{-3.3ex}

\section*{{\fontsize{11}{11}\textsf{\textbf{Acknowledgements\vspace{-1.2ex}}}}}
{\footnotesize \begin{spacing}{1.05}The authors thank R. Trivedi for insightful discussions. We gratefully acknowledge financial support from the Gordon and Betty Moore Foundation (GBMF4744). Additionally, K.Y. acknowledges the Nano- and Quantum Science and Engineering Postdoctoral Fellowship, D.V. acknowledges funding from FWO and the European Union Horizon 2020 Research and Innovation Program under the Marie Sklodowska-Curie grant agreement No 665501, R. T. acknowledges funding from the Kailath Graduate Fellowship. Part of this work was performed at the Stanford Nano Shared Facilities (SNSF)/Stanford Nanofabrication Facility (SNF), supported by the National Science Foundation under award ECCS-1542152.\end{spacing}
}

\vspace{-1.5ex}

\section*{{\fontsize{11}{11}\textsf{\textbf{Author contributions\vspace{-1.2ex}}}}}
{\footnotesize \begin{spacing}{1.05}N.V.S. performed and led the design, simulation, and fabrication of the accelerator. K.Y.Y. and Y.M. assisted with fabrication. D.V. assisted with design. K.L.L and D.S.B. conducted the electron acceleration experiment. R.J.E. performed the particle tracking simulations. L.S. provided the grating coupler design code. J.V., R.L.B., and O.S. organized the collaboration and supervised the experiments. All authors participated in the discussion and understanding of the results.\end{spacing}
}

\vspace{-1.5ex}

\section*{{\fontsize{11}{11}\textsf{\textbf{Data availability\vspace{-1.2ex}}}}}
{\footnotesize \begin{spacing}{1.05}Correspondence and requests for materials should be addressed to N.V.S.\end{spacing}
}

\vspace{-1.5ex}

\section*{{\fontsize{11}{11}\textsf{\textbf{Competing financial interests\vspace{-1.2ex}}}}}
{\footnotesize \begin{spacing}{1.05}The authors declare no competing financial interests.\end{spacing}
}
\clearpage

\onecolumn
\appendix 
\setcounter{figure}{0} 
\setcounter{equation}{0}
\onecolumn
\section* {Supplementary Information - On-chip integrated laser-driven particle accelerator}
% \vspace{-0.15 in}
\noindent Neil V. Sapra$^{*,\dagger,}$, Ki Youl Yang$^{\dagger}$, Dries Vercruysse$^{\dagger}$, Kenneth J. Leedle$^{\dagger}$, Dylan S. Black$^{\dagger}$, R. Joel England$^{\ddagger}$, \\Logan Su$^{\dagger}$, Yu Miao$^{\dagger}$, Olav Solgaard$^{\dagger}$, Robert L. Byer$^{\dagger}$, and Jelena Vu\u{c}kovi\'{c}$^{\dagger}$
\vspace{0.1 in}\\
\noindent
$^{*}$Corresponding author: nvsapra@stanford.edu \\
$^\dagger$E. L. Ginzton Laboratory, Stanford University, Stanford, CA, USA.\\
$^\ddagger$SLAC National Accelerator Laboratory, Menlo Park, CA, USA.

\section{Simulation analysis}
After obtaining an optimized accelerator design from the inverse design method, we then simulated a finite-length 30-period structure in a 3D finite-difference time-domain (FDTD) simulation. We compute the acceleration gradient spectrum through:

\begin{equation}
\label{eqn:freq_domain_gradient}
\begin{aligned}
G_z(E(\omega),\omega,\beta) = \frac{1}{L}\int_{0}^{L}e^{-i \omega z/\beta c_0}E_z(z,\omega)dz 
\end{aligned}
\end{equation}

\noindent By sweeping across different normalized electron velocities ($\beta = v/c$) and computing the acceleration spectrum for each $\beta$, we see in Figure 1 that we obtain optimal performance at a center wavelength of $\lambda = \SI{1.964}{\um}$ and electron velocity of $v=0.51c$. 

\begin{figure*}[h!]
  \centering
  \includegraphics[width=8.8cm]{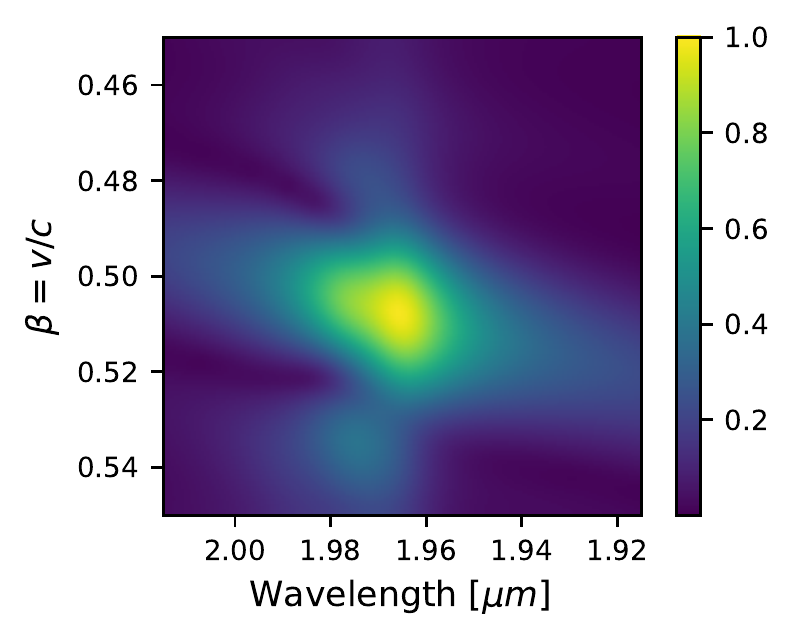}
  \vspace{-3ex}
  \caption{\textbf{Figure 1 \textbar \hspace{0.1pt} Frequency-domain acceleration gradient spectra.} Acceleration gradient spectra of a finite-length 30-period accelerator are computed at varied normalized electron velocities. Peak operating condition is determined to occur at roughly $\beta=0.51$ and $\lambda=\SI{1.964}{\um}$}
  \vspace{5ex}
  \label{si_figure:1}
\end{figure*}

After determining the peak operating wavelength for the 30-period accelerator to be $\SI{1.964}{\um}$ from the above analysis, we ran an additional FDTD simulation with a \SI{300}{fs} input pulse, centered at $\lambda=\SI{1.964}{\um}$ to match the pulse length of the laser used in the experiment. From the time-domain fields, we compute the accelerating gradients ($G_z$) and deflecting gradients ($G_y$) through:

\begin{equation}
\label{eqn:si_time_domain_gradient}
\begin{aligned}
G_k(t_0) = \frac{1}{L}\int_{0}^{L}{E_k(z,t_0+z/\beta c_0)}dz,
\end{aligned}
\end{equation}

\noindent where $t_0$ is the delay between the time of source injection and the electron entering the accelerator channel, and $L = \SI{30}{\um}$ is the length of the accelerator. To determine the optimal time delay, we evaluate the acceleration gradient, $G_z(t_0)$ as a function of the time delay, $t_0$, and plot the result in Figure 2a. In Figure 2b, we run a finer sweep of time delays from $t_0=\SI{750}{fs}$ to $t_0=\SI{770}{fs}$ and take the ratio of accelerating to deflecting gradients ($G_z/G_y$). We chose the peak at \SI{762.7}{fs} to calculate the gradients found in main text Figure 2b.

\begin{figure*}[h]
  \centering
  \includegraphics[width=8.8cm]{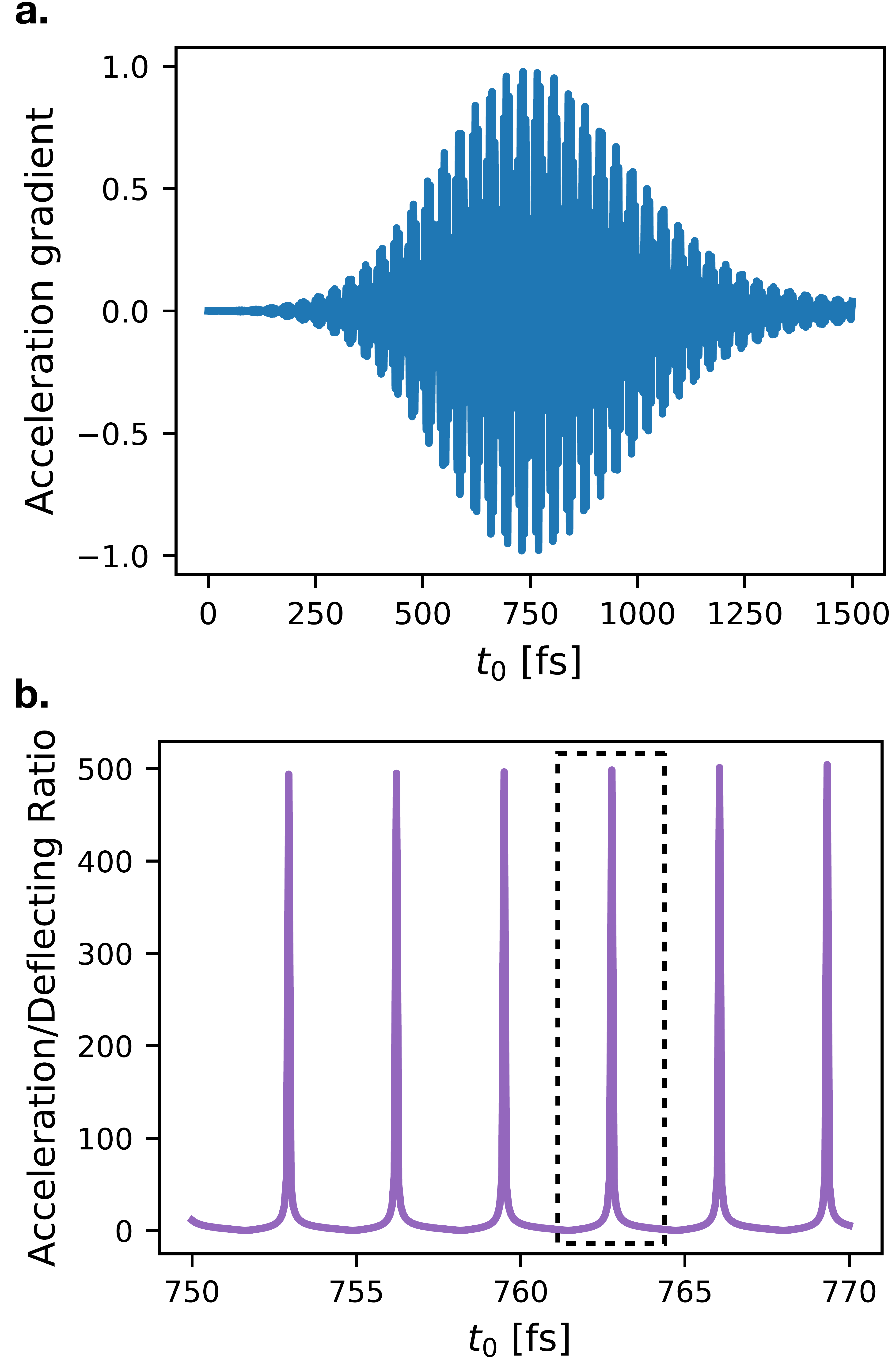}
  \vspace{-1ex}
 \caption{\textbf{Figure 2 \textbar \hspace{0.1pt} Determination of time delay for time-domain gradients.} \textbf{a.} Time-domain acceleration gradients (Equation 2) as a function of time-delay parameter, $t_0$. \textbf{b.} Finer sweep of time delay from $t_0=\SI{750}{fs}$ to $t_0=\SI{770}{fs}$ to compute the ratio of accelerating to deflecting gradients ($G_z/G_y$). Boxed peak represents the time delay value, $t_0=\SI{762.7}{fs}$, used in the main text for computing time-domain gradients.}
  \label{si_figure:2}
\end{figure*}

\section{Grating coupler design and performance}

The broadband grating couplers designed through inverse design were simulated in a 3D FDTD simulation (Lumerical FDTD). Mode expansion monitors were used to determine coupling into the TE0, TE2, and TE4 modes - as these modes are allowed by the symmetry of the input Gaussian source. In Figure 3 we see the coupling efficiency of the grating coupler into these modes. We indeed observe the desired broadband behavior and little coupling to higher-order modes. At \SI{1.964}{\um} the coupling efficiency into the TE0 mode is 14.3\%, TE2 mode is 0.57\% and TE4 mode is 0.18\%. We take the coupling into the remaining higher order modes to be zero. From these simulations, we then determine that 95\% of power input light is coupled into the fundamental TE0 mode.

\begin{figure*}[th!]
  \centering
  \vspace{2ex}
  \includegraphics[width=8.8cm]{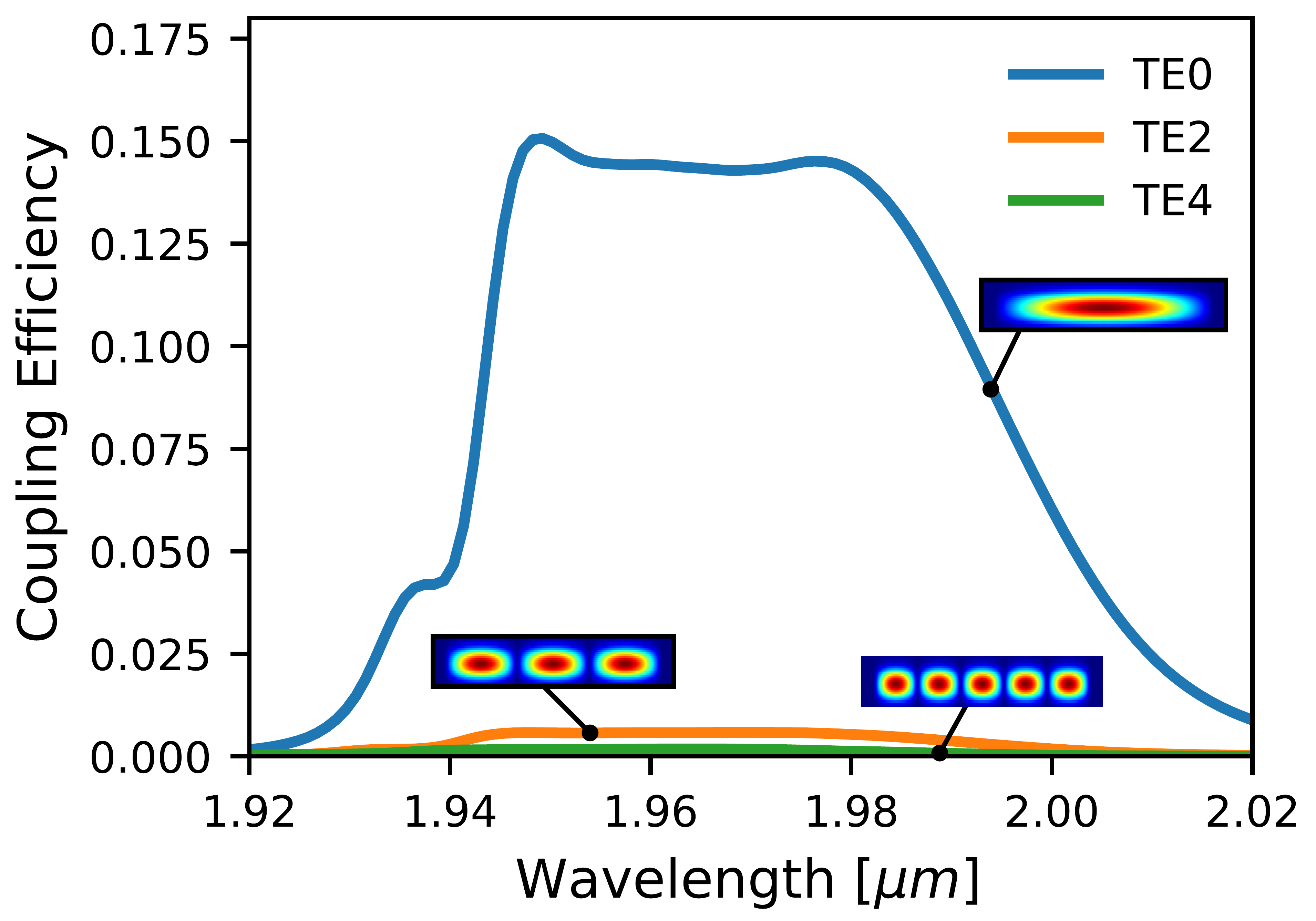}
  \vspace{-1ex}
  \caption{\textbf{Figure 3 \textbar \hspace{0.1pt} Coupling efficiency to waveguide modes through inverse designed grating coupler.} Coupling efficiency spectra of the inverse designed grating coupler in to TE0, TE2, and TE4 waveguide modes of a \SI{500}{nm} by \SI{30}{\um} waveguide.}
  \label{si_figure:3}
\end{figure*}

\section{Experimental electron energy spectra}

The electron energy spectra are obtained by averaging at least five frames from the micro-channel plate (MCP) detector, with each frame consisting of two seconds of integration. Below we include a randomly selected sample of five frames for each measurement. Figure 4 depicts the frames associated with the wavelength sweep (Figure 4b main text) and Figure 5 depicts the frames associated with the power sweep (Figure 4c main text). Error bars found in the main text were derived by computing the standard deviation of the shoulder energy modulation and maximum energy modulation of the collection of frames.

\begin{figure*}[th!]
  \vspace{0.655ex}
  \centering
  \includegraphics[width=17cm]{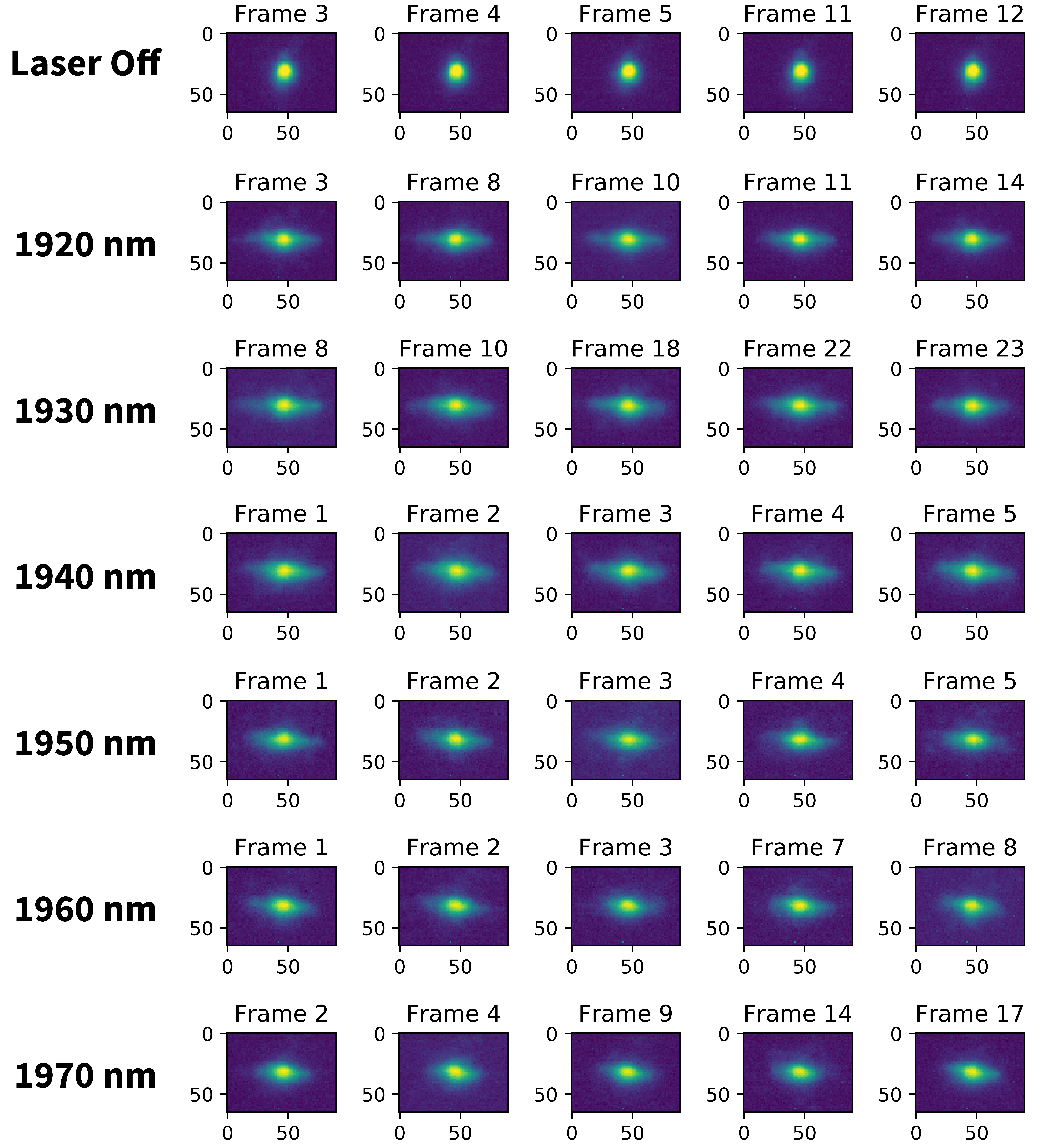}
  \caption{\textbf{Figure 4 \textbar \hspace{0.1pt} Wavelength sweep electron energy spectra frames.} Randomly selected sample of electron energy spectra frames at different center drive wavelength with constant incident power of \SI{2.75}{mW}.
  }
  \label{si_figure:4}
\end{figure*}

\begin{figure*}[th!]
  \vspace{0.655ex}
  \centering
  \includegraphics[width=17cm]{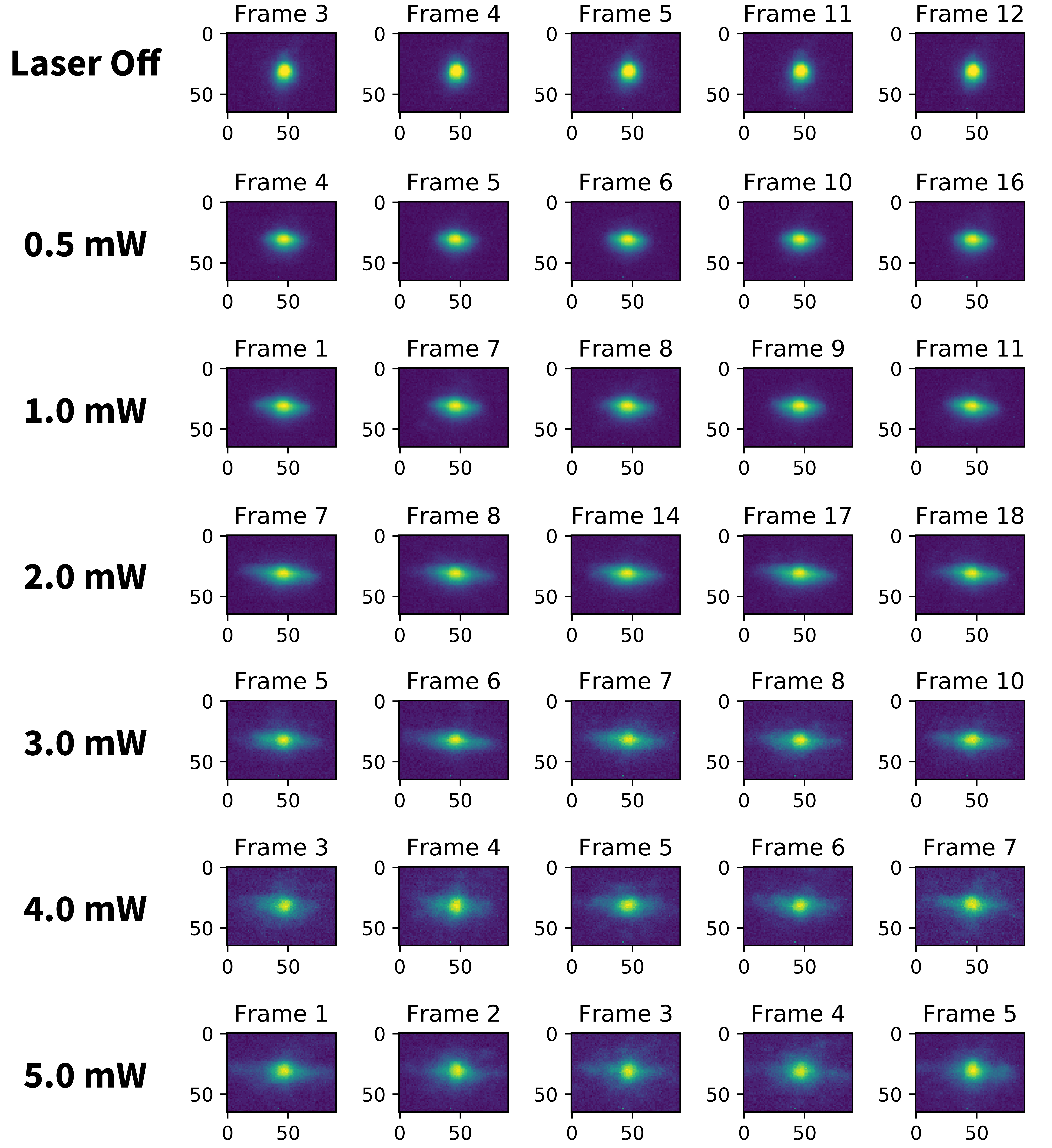}
  \caption{\textbf{Figure 5 \textbar \hspace{0.1pt} Power sweep electron energy spectra frames.} 
  Randomly selected sample of electron energy spectra frames at varying incident powers, but at constant center wavelength of $\lambda=\SI{1940}{nm}$}
  \label{si_figure:5}
\end{figure*}

% \begin{thebibliography}{1}
% \bibitem{Yariv} 
% Yariv, Amnon, and Pochi Yeh. "Photonics: optical electronics in modern communications (the oxford series in electrical and computer engineering)." Oxford University Press, Inc 231 (2006): 232.

% \bibitem{Vlasov} 
% Vlasov, Yurii A., and Sharee J. McNab. "Losses in single-mode silicon-on-insulator strip waveguides and bends." Optics express 12.8 (2004): 1622-1631.
 
% \end{thebibliography}

\end{document}